\documentclass[letterpaper,twocolumn,showpacs,pra]{revtex4} 
\usepackage{times}
\usepackage{graphics}
\usepackage{graphicx}
\usepackage{epsfig}
\usepackage{amsmath}
\begin{document}
\title{Large atom number Bose-Einstein Condensate machines}
\author{Erik W Streed$^1$, Ananth P Chikkatur$^2$, Todd L Gustavson$^3$, Micah Boyd$^1$, Yoshio Torii$^4$, Dominik Schneble$^5$, Gretchen K Campbell$^1$,David E Pritchard$^1$,Wolfgang Ketterle$^1$}




\affiliation{ $^1$ MIT-Harvard Center for Ultracold Atoms, Research
Laboratory of Electronics and Department of Physics, Massachusetts
Institute of Technology, Cambridge, MA 02139, USA \\
$^2$ Belfer Center for Science and International Affairs, 
Littauer P-14, 79 JFK St., Harvard University, Cambridge, MA 02138 USA \\
$^3$ Hansen Experimental Physics Laboratory End Station 3, Rm. M302, Department of Physics, Stanford University, Stanford, California, 94305 USA \\
$^4$ Institute of Physics, University of Tokyo, 3-8-1, Meguro-ku, Komaba, Tokyo 153-8902, Japan \\
$^5$ Physics A-106, Department of Physics and Astronomy, SUNY Stony Brook, Stony Brook, New York 11794, USA }






%

%
%
%

\date{\today}


\begin{abstract}
We describe experimental setups for producing large Bose-Einstein condensates of $^{23}$Na and $^{87}$Rb.
In both, a high flux thermal atomic beam is decelerated by a Zeeman slower and is captured and cooled in a magneto-optical trap. The atoms are then transfered into a cloverleaf style Ioffe-Pritchard magnetic trap and cooled to quantum degeneracy with radio frequency induced forced evaporation. Typical condensates contain 20 million atoms. The design includes a second vacuum chamber into which ultracold atoms can be transported with an optical tweezers. This allows the flexibility to rapidly prepare and perform a wide variety of experiments.
\end{abstract}
\maketitle

\tableofcontents

\section{Introduction}

It has been a decade since Bose Einstein condensation (BEC) in
atomic vapors was first observed \cite{Cornell-02, Ketterle-02}.
The transition from a classical thermal gas to the quantum
degenerate Bose-Einstein condensate occurs when the occupation
number of the lowest energy states (phase space density) $\rho=n
\lambda_{dB}^3$ is increased to $\sim 1$, where $n$ is the number
density and $\lambda_{dB}$ is the thermal de Broglie wavelength of
the atoms. So in principle, getting a BEC is easy:  you simply
cool down the gas until the critical phase space density is
reached. In practice, the procedure is more complicated.  A
variety of different techniques are needed to increase the phase
space density in several stages (Table \ref{tablepsd}).
Furthermore, each atom has different properties and requires 
modifications to the cooling techniques.  Major work by many groups
around the world has now extended the cooling techniques to an
impressive number of atomic species:  $^{87}$Rb
\cite{Anderson-95}, $^{23}$Na \cite{Davis-95}, $^{7}$Li
\cite{Bradley-95,Hulet-97}, $^{1}$H \cite{Kleppner-98}, $^{85}$Rb
\cite{Wieman-00}, $^{4}$He* \cite{Aspect-01,CohenTannoudji-01},
$^{41}$K \cite{Inguscio-01}, $^{133}$Cs \cite{Grimm-03},
$^{174}$Yb \cite{Takahashi-03}, and $^{52}$Cr
\cite{Griesmaier-05}.  Still, $^{23}$Na and $^{87}$Rb are the two
atoms which appear to have the most favorable properties for laser
and evaporative cooling and are the two work horses in the
field.

\renewcommand{\arraystretch}{1.3}
\begin{table}[ht!]
\begin{minipage}[ht!]{8 cm}
\begin{tabular}[t]{|ll|l|l|l|}
\hline
{\bf Stage}                    & {\bf n} (/cm$^3$) & {\bf Temperature}      & {$\mathbf{\rho}$}     \\\hline\hline
Oven                           & $10^{13}$         & $383$K                   & $10^{-14}$       \\\hline
Thermal Beam                   & $10^{7}$          & $v_{mp}=334$ m/s         & $10^{-20}$       \\\hline
Slowed Beam                    & $10^{7}$          & $v_{mp}=43$ m/s          & $10^{-18}$       \\\hline
Loading MOT\footnotemark[1]    & $10^{9}$          & $150\mu$K                & $10^{-8}$        \\\hline
Compressed MOT\footnotemark[1] & $10^{10}$         & $300\mu$K                & $4\times10^{-8}$ \\\hline
Molasses\footnotemark[1]       & $10^{10}$         & $10\mu$K                 & $6\times10^{-6}$ \\\hline
Magnetic trap                  & $10^{11}$         & $500\mu$K                & $2\times10^{-7}$ \\\hline
BEC Transition                 & $3\times10^{13}$  & $500$ nK                 & 2.61             \\\hline
Pure BEC                       & $10^{14}$         & ($250$ nK\footnotemark[2]) & (100)              \\\hline
\end{tabular}
\caption{Typical phase space densities ($\rho$) during BEC production in the $^{87}$Rb apparatus.}
\label{tablepsd}
\footnotetext[1]{Typical values, not measured separately}
\footnotetext[2]{Chemical potential}
\end{minipage}
\end{table}
\renewcommand{\arraystretch}{1}

A major difference between the various experiments is the way in which
atoms are laser cooled and then loaded into a magnetic trap (or
now sometimes into an optical trap) for evaporative cooling.  Our
approach at MIT employs atomic ovens and Zeeman slowing. Other
approaches use vapor cell magneto-optical traps, often in a double
MOT configuration and more recently surface MOTs.  An important figure of
merit of a BEC setup is the number of atoms in the condensate.
Large atom number allows better signal-to-noise ratios, greater
tolerance against misalignments, and greater robustness in
day-to-day operation.  Since 1996, the MIT sodium BEC setups have
featured the largest alkali condensates.  Our three setups
routinely produce condensates with atom numbers between 20 and 100
million.  Since diode lasers for rubidium are less expensive than
the dye lasers for sodium, most new groups have chosen to work in rubidium.
The most popular laser cooling setups for rubidium involve vapor 
cell MOTs, which do not obtain the condensate size of the MIT
sodium experiments. There has been a widespread perception in the
field that Zeeman slowing is the technique of choice for sodium
and vapor cell traps for rubidium. The construction of vapor cell MOT 
rubidium condensate machines is extensively detailed in the complementary
work of Ref. \cite{Lewandowski-03}.

When the Center of Ultracold Atoms was created at MIT and Harvard,
one major funded project was the translation of the techniques we
had developed for sodium to rubidium and to create a rubidium BEC
experiment with enormous condensates.  The successful
accomplishment of these goal is described in this paper. Since
we are in the unique situation to have two similar experimental
setups for Rb and Na, we are able to discuss similarities and
differences between the optimization for the two species.  Our conclusion is 
that the technique of using an atomic
beam and Zeeman slower works as well for rubidium as for sodium,
and we present the technical details of how to build an intense
slow beam for both atomic species.  Other figures-of-merit besides atom number
include simplicity and reliability.  In our experience,
the Zeeman slowing technique is by far the simplest technique to
generate an intense slow beam since it requires only a single
laser beam of modest power.  The length of the slower
and therefore the overall size of the vacuum setup may look
intimidating, but once it is built, it provides simple and
reliable day-to-day operation without need for realignment.

The third-generation sodium experiment and the rubidium experiment
were both designed with an additional vacuum chamber (``science
chamber'') into which the BEC or evaporatively cooled atoms can be
moved using optical tweezers.  The multi-chamber design allows us
to rapidly reconfigure the experimental setup in the science
chambers while keeping the BEC production chambers under vacuum.
This has allowed us to perform very different experiments in rapid
succession \cite{Gustavson-02, Leanhardt-02, Chikkatur-02A, Leanhardt-03B,
Leanhardt-03, Shin-04, Pasquini-04, Shin-04A, Shin-04B,Schneble-03,Schneble-04,Campbell-05}.

%
%
%
%
%

\section{System Overview}
\label{systemoverview}

\begin{figure}[h!]
\begin{center}
\epsfig{file=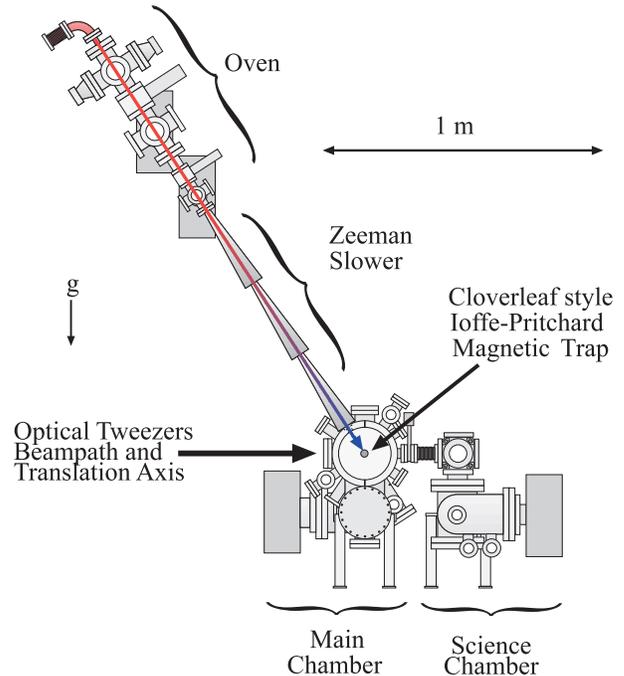, width=8 cm}
\caption{Vacuum system diagram and major subsystems. $^{87}$Rb apparatus shown.}
\label{figwholesystem}
\end{center}
\end{figure}

Fig. \ref{figwholesystem} illustrates the layout of our system. A thermal atomic beam emanates from the oven and is decelerated with the Zeeman slower. In the main chamber, the slowed atoms are captured and cooled with a six-beam magneto-optical trap (MOT) \cite{Raab-87}. To load the magnetic trap, the atoms are optically pumped into the F=1 hyperfine ground state. Atoms in the F=1, m$_F$=-1 state are held by their attraction to the field minimum of the Ioffe-Pritchard magnetic trap.

The trapped sample is evaporatively cooled by removing hotter atoms through radio frequency (RF) induced transitions into untrapped states. Reducing the RF frequency lowers the effective depth of the magnetic trap, allowing us to progressively cool to higher densities and lower temperatures until the atoms reach BEC. Magnetically trapped atoms in the F=2, m$_F$=+2 state have also been evaporated down to BEC.

Ultracold atoms can be transported from the main chamber into an adjoining auxiliary ``science chamber'' by loading the atoms into the focus of an optical tweezer and then translating the focus. In this manner we have transported $^{23}$Na BECs \cite{Gustavson-02}. Vibrational issues during transport cited in \cite{Gustavson-02} were reduced by the use of Aerotech ABL2000 series air bearing translation stages.

Technical issues related to the greater mass and higher three body recombination rate in $^{87}$Rb were overcome by transporting ultracold atoms just above the transition temperature $T_c$, and then evaporating to BEC at the destination. The oven and Zeeman slower are tilted by 57$^\circ$ from horizontal to allow a horizontal orientation for the weak trapping axes of both the optical tweezers and magnetic trap.

Trapping ultracold atoms requires that they be isolated from the surrounding environment. The laser and magnetic trapping techniques confine the atoms in the center of the chamber, out of contact with the room temperature chamber walls. The atoms are still exposed to thermal black body radiation from the chamber walls, but are transparent to most of the spectrum. The transitions which the black body radiation can couple to are the optical transitions used for laser cooling and the microwave hyperfine transitions. For optical transitions, which have energies much greater then $k_B T$ the excitation rate is $ \frac{3}{\tau_{optical}} \exp{\left( -\hbar\omega_{optical} / k_B T \right) }$, where $\omega_{optical}$ is the frequency of the transition and $\tau_{optical}$ the lifetime of the excited state. For rubidium in a 25$^\circ$C chamber this gives a characteristic excitation lifetime of $\sim$56 billion years. Raising the chamber temperature to 680$^\circ$C increases the optical excitation rate into the experimentally relevant domain of once per minute. The hyperfine transitions are significantly lower in energy compared to $k_B T$ and have an excitation rate of $\frac{3}{\tau_{hfs}}\frac{k_B T }{ \hbar \omega_{hfs} }$, which is once per year at 25$^\circ$C in $^{87}$Rb since the ground state hyperfine spontaneous decay lifetime $\tau_{hfs}$ in an alkali atom is thousands of years. Neither of these excitation rates are limitations on current experiments.

\begin{figure}[ht!]
\begin{center}
\epsfig{file=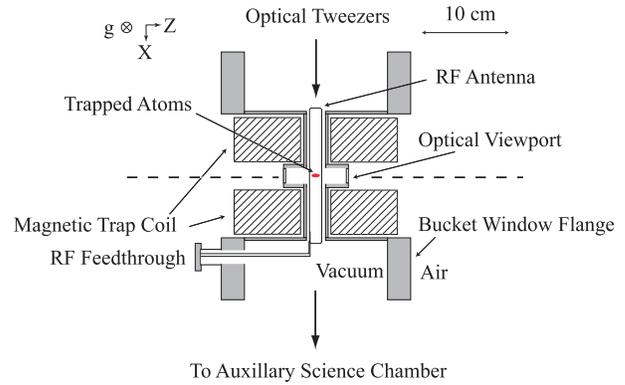, width=8 cm}
\caption{ Main chamber cross section showing re-entrant bucket windows, magnetic trap coils, and RF antenna. View from above.}
\label{figbucketwindows}
\end{center}
\end{figure}

Collisions with background gas molecules result in loss from the trap, necessitating low vacuum pressure for long atom cloud lifetime. We can magnetically trap ultracold atoms with lifetimes of several minutes in the $<10^{-11}$ torr ultrahigh vacuum (UHV) environment of the main production chamber. To achieve this vacuum performance we have followed the general guidelines set out in Ref. \cite{VacuumTechnologyUsersGuide} for constructing vacuum systems. The main chamber body was constructed of nonmagnetic 304 stainless steel and then electropolished to reduce the surface roughness. The only component placed inside the chamber was the RF evaporation antenna coil (Fig. \ref{figbucketwindows}).

The cloverleaf-style Ioffe-Pritchard magnetic trap coils fit inside two re-entrant bucket windows \footnote{Simon Hanks of UKAEA, D4/05 Culham Science Center, Abingdon,UK}, allowing them to be outside the chamber with an inter coil spacing of 25 mm (Fig. \ref{figbucketwindows}). The Zeeman slower tube is mounted between the main chamber and the oven chamber. The Zeeman slower coils that are around the Zeeman slower tube are also outside of the vacuum system but cannot be removed without breaking vacuum.

After assembling the chamber, we pumped out the system and reached UHV conditions by heating the system to accelerate outgassing. We heated the main chamber to 230$^\circ$C and the Zeeman slower to 170$^\circ$C (limited by the coil epoxy). Using a residual gas analyzer to monitor the main chamber, we ``baked'' until the partial pressure of hydrogen was reduced to less than  $10^{-7}$ torr and was at least ten times greater than the partial pressure of other gases. A typical bakeout lasted between 3 and 9 days, with temperature changes limited to less than $50^\circ$C/hour. While we acknowledge the merit of using dry pumps as recommended in Ref. \cite{Lewandowski-03}, we have not had any detrimental experiences using oil sealed rotary vane roughing pumps to back our turbo pumps. The vacuum in the main chamber is preserved after bakeout with a 75 L/s ion pump and a titanium sublimation pump. Refer to Sec. 3.4 of Ref. \cite{Chikkatur-02} for more details of our bakeout procedures.

%
%
%
%

\section{Oven}
\label{Rboven}

\begin{figure*}[ht!]
\epsfig{file=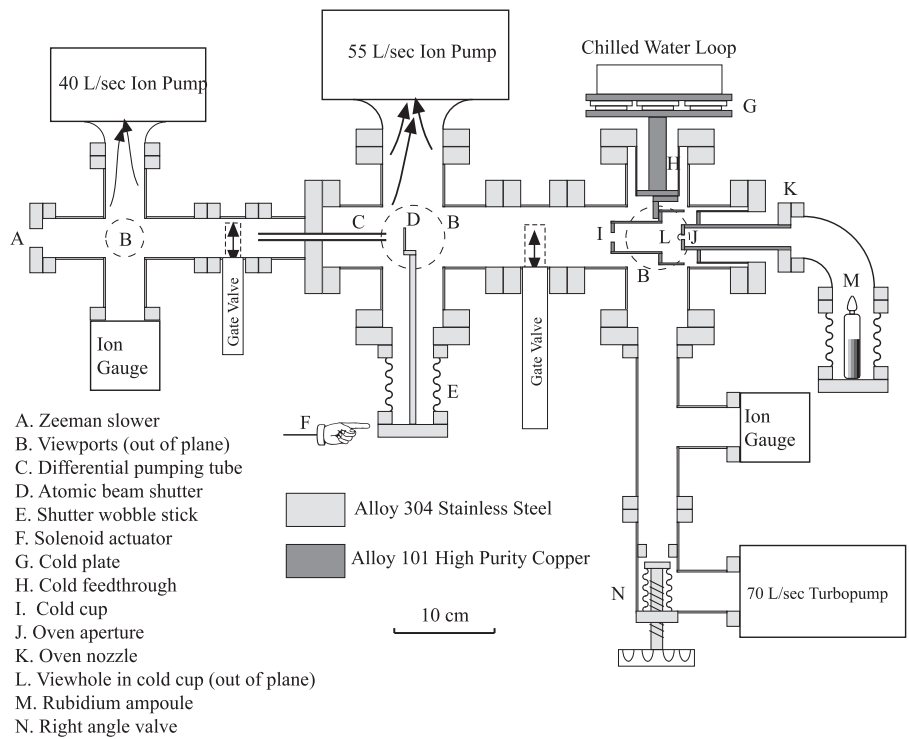, width=12 cm}
\caption{Effusive rubidium beam oven. Rubidium metal (M) is heated to between 110$^\circ$C and 150$^\circ$C, creating a $p_{Rb} \sim0.5$ millitorr vapor which escapes through a 5 mm diameter hole (J). A 7.1mm diameter hole in the cold cup (I), 70 mm from the nozzle, allows 0.3\% of the emitted flux to pass through, forming an atomic beam with a flux of $\sim10^{11}$ atoms/s. The remainder is mostly (99.3\%) captured on the -30$^\circ$C, $p_{Rb} \approx2.5\times10^{-10}$ torr, surface of the cold cup. We chop this beam with a paddle (D) mounted to a flexible bellows (E). The differential pumping tube (C) and Zeeman slower tube (A) consecutively provide 170x and 620x of pressure isolation between the oven and main chambers.}
\label{figovendiagram}
\end{figure*}

We generate large fluxes of thermal atoms for Zeeman slowing from effusive atomic beam ovens. An effusive beam is created by atoms escaping through a small hole in a heated chamber \cite{AtomicMolecularBeamMethods}. The higher vapor pressure of rubidium requires a more complicated design, but lower operating temperature (110-150$^\circ$C Rb, 260$^\circ$C Na.) At room temperature, the vapor pressure of sodium ($\approx 2\times10^{-11}$ torr \cite{Alcock-84}) is roughly compatible with our UHV main chamber environment, while that of rubidium ($\approx 4\times10^{-7}$ torr \cite{Alcock-84}) is not. This dictated that the design of the rubidium oven prevent contaminating the main chamber with rubidium. Because of its greater complexity, further discussion will focus on the rubidium oven (Fig. \ref{figovendiagram}). We expect that the rubidium oven would work as well for sodium, but instead we used a simpler design described in Ref. \cite{Chikkatur-02}.

A combination of active pumping and passive geometrical techniques were used to reduce extraneous rubidium transfer to the main chamber. A cold cup (I) is used to reduce rubidium vapor in the oven chamber by almost completely surrounding the oven aperture (J) with a cold surface. After bakeout, the combination of cold cup and oven chamber ion pump has achieved pressures as low as $\sim10^{-9}$ torr, although we have successfully made BECs with pressures of up to $\sim10^{-6}$ torr in this region. A combination of a differential pumping tube, ion pump, and the Zeeman slower tube provides a pressure differential of over 3 orders of magnitude between the oven and main chamber. This is sufficient to isolate the UHV environment from an oven pressure dominated by rubidium vapor at room temperature. When the oven is opened to replace rubidium and clean the cold cup, the main chamber vacuum is isolated with a pneumatic gate valve. The second gate valve can be used in case of failure of
 the first. While not used in our system, designers may want to consider gate valves with an embedded window from VAT to allow optical access along Zeeman slower or tweezer beamlines during servicing.

The oven is loaded with a sealed glass ampoule containing 5 g of rubidium in an argon atmosphere. To add rubidium, the ampoule is cleaned, placed in the oven, and baked out under vacuum while still sealed. We then break the ampoule under vacuum and heat the oven to 110$^\circ$C to produce the atomic beam. During operation, the machine is run as a sealed system, without the turbo-mechanical pump, to prevent accidental loss of the main chamber vacuum. Oven temperatures from 150$^\circ$C down to 110$^\circ$C produced similar sized $^{87}$Rb BECs. Reducing the oven temperature increased the time between rubidium changes to greater than $1000$ hrs of operating time. This long operating cycle precluded the need for more complex recycling oven designs \cite{Walkiewicz-00}.

%
%
%
%


\section{Zeeman slower}
\label{slower}

The atomic beams are slowed from thermal velocities by nearly an order of magnitude by scattering photons from a resonant, counter-propagating laser beam. When a photon with momentum $\hbar k$ ($k=2\pi/\lambda$) is absorbed or emitted by an atom with mass $m$, the atom will recoil with a velocity change of $v_{r}= ~\hbar k/m$ to conserve momentum. Atoms can resonantly scatter photons up to a maximum rate of $\Gamma/2$, where $1/\Gamma=\tau$ is the excited-state lifetime. This results in a maximum acceleration $a_{max}= \frac{\hbar k \Gamma}{2 m}$ ($1.1\times10^5 \mbox{m/s}^2$ Rb, $9.3\times10^5 \mbox{m/s}^2$ Na). As the atoms decelerate, the reduced Doppler shift is compensated by tuning of the Zeeman shift with a magnetic field \cite{Phillips-82} to keep the optical transition on resonance. We designed our slower to decelerate the atoms at a reduced rate $f a_{max}$ where $f \sim 50\%$ is a safety factor to allow for magnetic field imperfections and finite slower laser intensity.

\begin{figure}[ht!]
\begin{center}
\epsfig{file=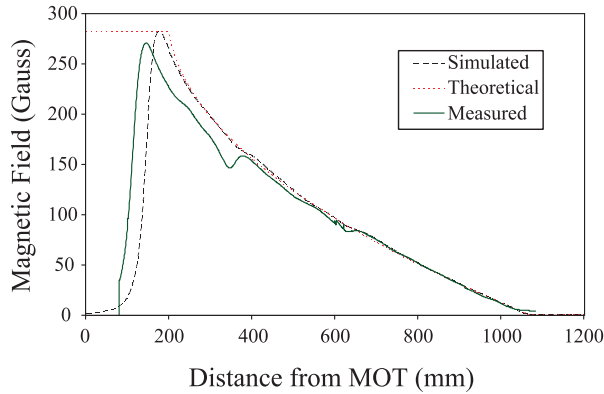, width=8 cm}
\caption{ Magnetic field profile of the rubidium Zeeman slower, not including uniform bias field. The theoretical line shows the desired magnetic field profile for atoms decelerated from 330 m/s to 20 m/s at 60\% of the maximum intensity limited deceleration (f=53\% of $a_{max}$.) The simulated line depicts the expected field from slower coils with the winding pattern in Fig. \ref{figslowerwinding} of Appendix \ref{slowerappendix}. Prominent bumps in the measured field were subsequently smoothed with additional current carrying loops.}
\label{figfieldprof}
\end{center}
\end{figure}

Our slowers are designed along the lines of Ref. \cite{Barrett-91}, with an increasing magnetic field and $\sigma^-$ polarized light scattering off the F=2, m$_{F}$=-2 $\rightarrow$ F$^\prime$=3, m$_{F^\prime}$=-3 cycling transition. Before the slowing begins, the atoms are optically pumped into the F=2,m$_{F}$=-2 state. The large magnetic field at the end of the slower corresponds to a large detuning from the low velocity, low magnetic field resonance frequency. This large detuning allows the slowing light to pass through the MOT without distorting it due to radiation pressure. Within the slower, the quantization axis is well-defined by the longitudinal magnetic field and optical pumping out of the cycling transition is strongly suppressed by the combination of light polarization and Zeeman splitting.

We slow $^{87}$Rb atoms from an initial velocity of $\sim$350 m/s with a tailored 271 G change in magnetic field (Fig. \ref{figfieldprof}). An additional uniform $\sim$200 G bias field was applied along the length of the slower to ensure that neighboring hyperfine levels were not near resonance in either the slower or the MOT. The slower cycling transition light is detuned -687 MHz from the F=2 $\rightarrow$ F$^\prime$=3 transition. The slowing laser intensity is $I/I_{sat} \approx 8$, giving a maximum theoretical deceleration of 89\% of $a_{max}$. ``Slower repumping'' light copropagates with the cycling transition light and is detuned -420 MHz from the F=1 $\rightarrow$ F$^\prime$=1 transition to match the Doppler shift of the unslowed thermal atoms from the oven. A flux of $\sim10^{11}$ $^{87}$Rb atoms/s with a peak velocity of 43 m/s was measured from our slower with an oven temperature of 150$^\circ$C. This is signifigantly greater flux then the $8\times10^8$ Rb/sec vapor cell loading rate quoted by \cite{Lewandowski-03}.

\begin{figure}[ht!]
\begin{center}
\epsfig{file=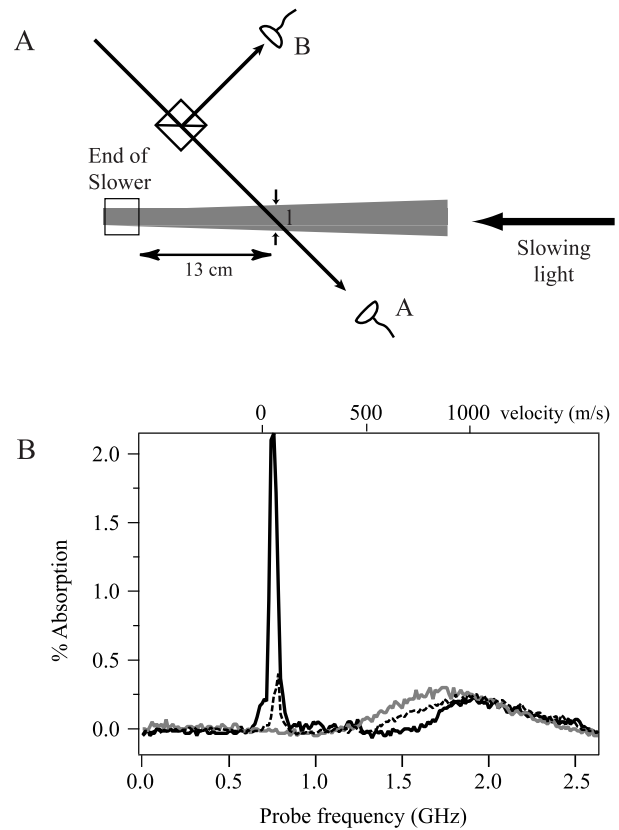, width=8 cm}
\caption{ Sodium slower performance. A. Schematic of the differential absorption measurement of the slowed atomic beam. The slowed atomic beam is shown in the gray. The photodiode signals $A$ \& $B$ are subtracted and amplified. B. Typical absorption signal for $^{23}$Na beam at 45$^\circ$ probe angle. The black solid line is the slowed beam with both slower solenoids fully energized. The dashed line is with only the increasing field slower solenoid, and the gray line is the raw  atomic beam without any slowing. The top scale converts the probe frequency into a velocity scale relative to the F = 2 cycling transition.}
\label{figNaslowertesting}
\end{center}
\end{figure}

The higher temperature of the sodium oven, along with the atoms' lower mass,
results in a greater initial velocity of 950 m/s. This requires a slower with a 
much larger magnetic field change of 1150 Gauss. To reduce the maximum magnitude 
of the magnetic fields we use the spin flip variant of the increasing field design 
by shifting the zero crossing of the magnetic field from the beginning of the slower 
to the middle. The first segment then becomes a decreasing field slower, with current
flowing in the opposite direction of the second, increasing field segment. In the low
magnetic field region near the zero crossing, the atoms are theoretically vulnerable 
to optical pumping out of the cycling transition, but we did not find this to be a 
problem experimentally. Ref. \cite{Slowe-04} has demonstrated a high flux spin flip 
style slower for $^{87}$Rb. The sodium slowing beam is detuned -1.0 GHz from the F=2 
$\rightarrow$ F$^\prime$=3 transition and has an intensity of $I/I_{sat}\approx 4$, 
giving a laser power limited maximum deceleration of 80\% of $a_{max}$. Unlike the 
rubidium slower, light for optical pumping was generated by adding 1.75 GHz sidebands
 to the slowing light using an electro-optical modulator.

The sodium slower coils were broken up such that the first segment had an initial field of 440 G and a length of 52 cm and the second segment had a final field of 710 G and a length of 43 cm. The sodium slower was tested as depicted in Fig. \ref{figNaslowertesting}, with a measured flux of $3\times10^{11}$ $^{23}$Na atoms/s with a peak velocity of 100 m/s.

%
%
%
%

\section{Lasers}
\label{lasers}

\begin{figure}[ht!]
\begin{center}
\epsfig{file=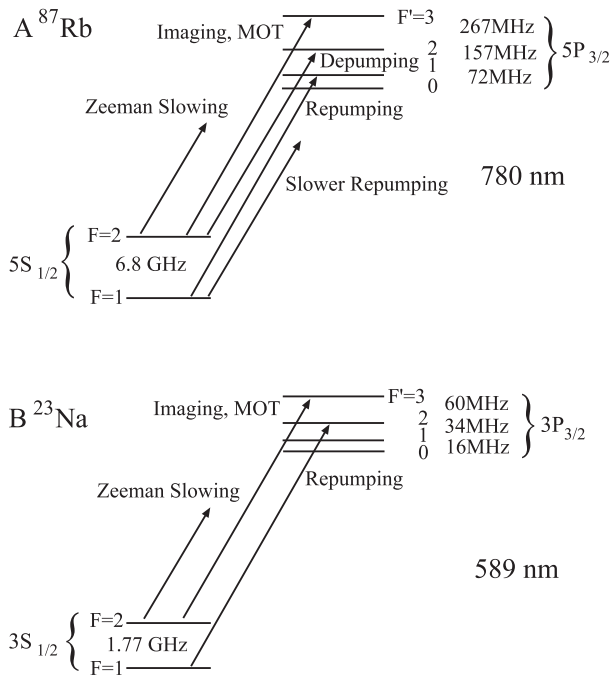, width=8 cm}
\caption{Simplified level structure of $^{87}$Rb (A) and $^{23}$Na (B) with relevant transitions, hyperfine splittings, and laser frequencies.}
\label{leveldiagrams}
\end{center}
\end{figure}

Resonant laser light is used to slow, cool, trap, and detect the atoms. All laser light is prepared on a separate optics table and delivered to the apparatus (Fig. \ref{figwholesystem}) through single-mode optical fibers. Because stray resonant light can heat the atoms during evaporation, black cloth separates the two tables. All frequency shifting and attenuation of the light is done with acousto-optic modulators. Mechanical shutters are also placed in front of each fiber coupler to block any light which might leak through the modulators and disturb the atoms. Atomic energy levels and laser frequencies used are indicated in Fig. \ref{leveldiagrams}.

We use different techniques for generating laser light at the resonant wavelengths of $^{87}$Rb (780 nm) and $^{23}$Na (589 nm). Commercially available (Toptica DL100,TA100) external cavity diode lasers and semiconductor tapered amplifiers are used to create 350 mW and 35 mW of light resonant with the $^{87}$Rb F=2$\rightarrow$ F$^{\prime}$=3 and F=1$\rightarrow$ F$^{\prime}$=1 transitions at 780 nm respectively. The lasers are stabilized with a polarization sensitive saturated absorption spectroscopy lock \cite{Yoshikawa-03,Pearman-02}. This modulation-free technique optically creates a derivative signal of the absorption spectra that is locked with a proportional+integral gain servo loop. The locking signal fluctuation indicates a frequency jitter of $<1$MHz over several seconds, which is much less than the 6.1 MHz natural linewidth of $^{87}$Rb.

The $^{87}$Rb MOT uses a total of 60 mW of light near the F=2$\rightarrow$F$^\prime$=3 cycling transition for trapping/cooling. The F=2$\rightarrow$F$^{\prime}$=3 transition in the MOT is only approximately a closed cycle and atoms are often optically pumped into the F=1 ground state. To ``repump'' these atoms back into the F=2 state we use 10 mW of light on the F=1$\rightarrow$ F$^{\prime}$=1 transition. Likewise, to deliberately transfer atoms from the F=2 to F=1 manifold we can introduce a few mW of ``depumping'' light resonant with the F=2$\rightarrow$ F$^{\prime}$=2 transition. Powers are quoted after fiber coupling, measured as delivered to the apparatus table. After frequency shifting the slower cycling and repumping light to their desired detunings only a few mW of power are available. Each of these beams is then amplified to 35-40 mW by injection locking \cite{SiegmanLasers} a free running Sanyo DL7140-201 laser diode. The two amplified beams are then overlapped and coupled into a fiber, which delivers 18 mW of slower cycling light and 6 mW of slower repumping light.

For $^{23}$Na we use a Coherent 899 dye laser pumped by a Spectra Physics Millenia laser (532 nm, $8.5\,{\rm W}$). Typically 1.2 W of 589 nm light is generated by the dye laser. The laser frequency was referenced to an external saturation-absorption lock-in scheme and locked to a Fabry-Perot cavity. Stable operation was improved by using a precision dye nozzle (Radiant Dyes, Germany), high pressure dye circulator at 12 bars, and stabilized temperatures for the room and dye.

For more detailed information on the generation of the laser light for sodium MOTs, see Sec.~$3.4$ of Ref. \cite{Stamper-Kurn-00}. Typical delivered laser powers are 80 mW for the MOT light, 20 mW for the repumping light, 40 mW for the slowing light and less than one mW for the imaging beam. Electro-optic modulators allow the addition of high frequency sidebands ($\sim$1.8 GHz) on the slowing and MOT light for repumping without the use of an additional laser beam. Recent advances in single frequency high power fiber and diode pumped solid state lasers \footnote{IPG Photonics EAD and RLM series amplifiers.} have made nonlinear techniques such as sum frequency generation \cite{Moosmuller-97, Bienfang-03} and frequency doubling \cite{Thompson-03} interesting alternatives as resonant light sources.

\section{Magneto Optical Trap}
\label{mot}
The MOT \cite{Raab-87} is the workhorse of atomic physics for creating large samples of ultracold atoms. We use a six-beam MOT, which doubles as an optical molasses when the magnetic gradient field is off. Similar to Ref. \cite{Lewandowski-03} the $^{87}$Rb apparatus uses a bright MOT. The $^{87}$Rb MOT equilibrates to around $4\times10^{10}$ atoms after $\sim2$ s of loading, operating in a magnetic field gradient of 16.5 G/cm with cycling beams detuned -18 MHz from the F=2$\rightarrow$ F$^{\prime}$=3 transition and a peak intensity 5.3 mW/cm$^2$. To increase the efficiency of the transfer into the magnetic trap, we briefly compress the $^{87}$Rb MOT and then switch off the magnetic field gradient to cool the atoms with optical molasses. The $^{87}$Rb MOT is compressed by linearly ramping the gradient to 71 G/cm in 200 ms and simultaneously sweeping the detuning to -45 MHz in 400 ms. We use 5ms of ``gray'' molasses, where the repumper power is dropped by 95\%, the optical trapping power is ramped down to 50\%, and the detuning is swept from -18MHz to -26MHz. The molasses phase requires reduction of imbalances in intensity between beams and also to residual magnetic fields \cite{Shang-91}. After the molasses phase, 0.5-1 ms of ``depumping'' light is applied to put all the $^{87}$Rb atoms into the F=1 level before loading into the magnetic trap. Exact MOT and molasses parameters were found through empirical optimization, and all listed numbers should be considered as guides.

The $^{23}$Na apparatus uses a dark-spot MOT \cite{Ketterle-93}, with a detuning of -15MHz, peak beam intensity of 8.8 mW/cm$^2$ and a magnetic field gradient of 11G/cm. A 4 mm diameter opaque circle blocks light in the middle of a single repumper beam, creating a region at the center of the MOT where trapped atoms are optically pumped into the F=1 state. The $^{23}$Na MOT equilibrates after a few seconds of loading. The effectiveness of the dark-spot in $^{23}$Na has precluded the need for the compression and molasses phases as in $^{87}$Rb. 99\% of trapped atoms reside inside the dark spot, and the $^{23}$Na atoms are sufficiently cold and dense to be directly loaded into the magnetic trap.

\section{Magnetic Trap}
\label{magtrap}

\begin{figure}[ht!]
\begin{center}
\epsfig{file=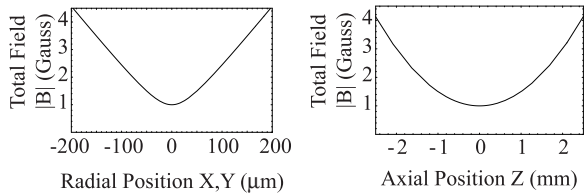, width=8 cm}
\caption{Profile of the Ioffe-Pritchard trap magnetic field magnitude. The trap parameters are $B^\prime$=223 G/cm, $B^{\prime\prime}$=100 G/cm$^2$, and $B_0$=1G.}
\label{figiptrapprofile}
\end{center}
\end{figure}

Atoms in weak magnetic field seeking states can be trapped in a magnetic field minimum. Our magnetic trap is a high current Ioffe-Pritchard (IP) trap with a cloverleaf style winding that can hold F=1, m$_F= -1$ or F=2, m$_F=+2$ ground state atoms of $^{87}$Rb and $^{23}$Na with long lifetimes. An IP trap has an anisotropic, ``cigar''-shaped, 3D harmonic trap for energies which are small compared to the trap minimum $g_F m_F \mu_B B_0$ and a 2D linear/1D harmonic trap at higher energies (See Fig. \ref{figiptrapprofile} and Appendix \ref{appendixMagtrap}). This linear regime at higher energies (higher cloud temperatures) is more efficient for evaporatively cooling hot atoms \cite{Ketterle-96a}, while the finite bias field at the minimum prevents Majorana spin flip loss of colder atoms. 

\begin{figure}[ht!]
\begin{center}
\epsfig{file=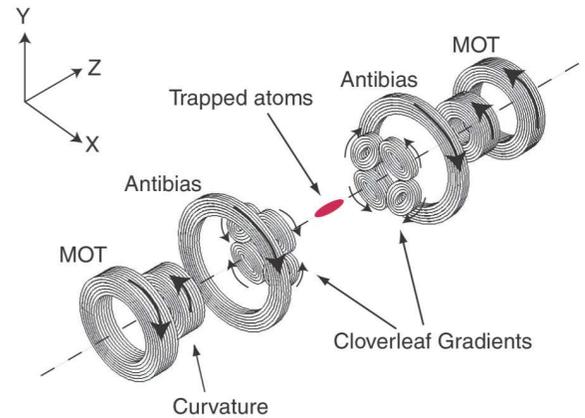, width=8 cm}
\caption{Exploded view of the cloverleaf style Ioffe-Pritchard trap coils. Arrows indicate the direction of current flow.
MOT coils are not on during magnetic trapping. Performance and design details are listed in Table \ref{tablecoilvalues} of Appendix \ref{appendixMagtrap}}
\label{figexplodedcoils}
\end{center}
\end{figure}

Fig. \ref{figexplodedcoils} shows an expanded view of the magnetic trap coils. The two sets of four cloverleaf coils create radial gradients $B'$ along $\hat{x}$ and $\hat{y}$, while the curvature coils produce a parabolic field curvature $B''$ in the $\hat{z}$ direction. The curvature coils also produce a substantial bias field (Table \ref{tablecoilvalues}, Appendix \ref{appendixMagtrap}) along $\hat{z}$, which is balanced by a roughly homogeneous field from the antibias coils, resulting in a low residual bias field $B_0$ of  $\sim $1 G at the center of the trap. The subtraction of the large magnetic fields from the curvature and antibias coils can make the residual bias field $B_0$ susceptible to jitter from current noise. To prevent this we drive current through both coils in series from the same power supply (Appendix \ref{appendixMagtrap}, Fig. \ref{figaxialcircuit}), reducing the effect of current noise in the residual bias field $B_0$ by $\approx$30. When assembled the anti
bias coils enclose the cloverleaf coils and the MOT coils surround the curvature coils.

When the magnetic trap is initially turned on, the strength and shape of the confinement are adjusted to match that of the laser cooled atoms to preserve phase space density. Additional current is applied to the curvature coils, increasing the residual bias field and decreasing the radial confinement to make a roughly spherical magnetic trap that match the spherical MOT. After loading atoms in the trap, the additional curvature coil current is reduced over one second to adiabatically change the trap geometry to the tightly confining cigar shape, favorable for evaporative cooling. Sec. 2.3.2 of Ref. \cite{Varenna} has an extensive discussion of mode matching magnetic traps to MOTs. The adiabatic compression technique is reviewed in Ref. \cite{Ketterle-96a}.

\section{Control and Imaging}
\label{sequence}

Two computers run the apparatus; one controls the various parts of experiment and the other processes images from a camera which images the atoms. The control computer has custom built National Instruments (NI) LabWindows based software to drive analog (2 NI Model PCI 6713, 8 channels of 12 bit analog, 1MS/s update) and digital output (2 NI  Model PCI-6533, 32 channels of binary TTL, 13.3 MS/s update) boards. The control computer also controls an Agilent 33250A 80MHz function generator through a GPIB interface, and triggers a Princeton Instruments NTE/CCD-1024-ED camera through a ST-133 controller to capture the absorption images.

BECs are typically imaged 10-40 ms after release from the trap. Ref. \cite{Castin-96,Varenna} provides the details of analyzing condensates after free expansion. Atoms are first optically pumped into the F=2 state in $200\mu$s and then an absorption image is taken using on resonance F=2 $\rightarrow$ F$^{\prime}$=3 light. Detuning off resonance causes dispersion (lensing) as the light passes through the cloud of atoms and can distort the image. The intensity of the imaging probe is kept lower than the saturation intensity to prevent bleaching of the transition, which would lead to errors in atom number counting. Typical exposure times are 50-200 $\mu$s. Sec. 3 of Ref. \cite{Varenna} discusses other imaging techniques that can also be used to probe BECs.

\section{Evaporation}
\label{magtrapbec}
Evaporative cooling works by selectively removing hot atoms from the trapped cloud, while the remaining atoms rethermalize to a lower temperature. The efficiency of cooling depends on $\eta$, the ratio of trap depth or energy of the escaping atoms to the temperature $k_B T$, and is reduced by the rate of heating. The speed of this process depends on how quickly the atoms rethermalize. In a magnetic trap evaporation is implemented through RF induced transitions between trapped and untrapped states. A given RF frequency corresponds to a shell of constant $\mu_m\left|\mathbf{B}\right|$ where the transitions occur. Atoms that pass through this shell enter untrapped states and are lost; thus RF provides a flexible mechanism to control the magnetic trap depth. Our RF antenna consists of two rectangular loops of wire, 10 cm x 2 cm, positioned 3 cm above and below the condensate as depicted in Fig. \ref{figbucketwindows}.

\begin{figure}[ht!]
\begin{center}
\epsfig{file=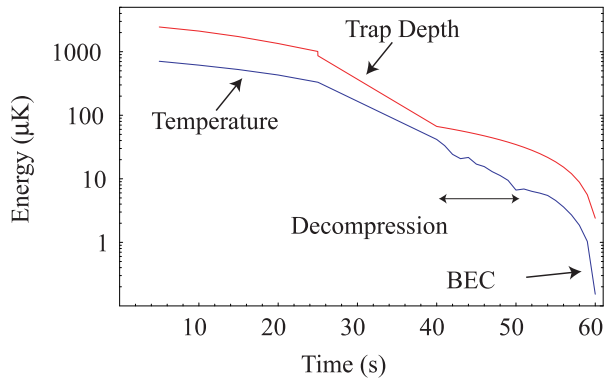, width=8 cm}
\caption{Typical temperature and trap depth during evaporation to BEC in $^{87}$Rb. The trap is decompressed between t=40-50 s by changing the trap parameters $B^\prime$= 223 $\rightarrow$ 54 G/cm, $B^{\prime\prime}$=99 $\rightarrow$ 25 G/cm$^2$, and $B_0$=1.4 $\rightarrow$ 0.87 G.}
\label{figevapplot}
\end{center}
\end{figure}

To evaporate thermal atoms to a BEC, we sweep the RF frequency over several seconds using an Agilent 33250A synthesizer amplified with a 5 W RF amplifier (Mini-Circuits ZHL-5W-1).
Typical evaporation curves for $^{87}$Rb would ramp from 60 MHz down to $\sim$0.8 MHz in 15 to 40 seconds. Forced RF evaporative cooling is very efficient, increasing phase space density by $>10^6$ (Table \ref{tablepsd}). Fig. \ref{figevapplot} shows the drop in temperature as the trap depth (calculated from the RF frequency) is lowered during evaporation of $^{87}$Rb. Evaporation curves are frequently adjusted in the interest of tuning evaporation speed, atom number, density, and/or reproducibility. For instance, the atom number can be increased by decompressing the magnetic trap near the end of the evaporation. This reduces the effects of three body recombination heating by lowering the final condensate density. Such decompression techniques have allowed us to create nearly pure condensates with $N_c \approx 20\times10^6$ in both $^{87}$Rb and $^{23}$Na with lifetimes in excess of 5 seconds.

Decompressing the trap shifts its center due to gravitational sag and imperfections in the balance of magnetic fields between the coils. Such movements can excite oscillations in the cloud, which results in the condensation of BECs which are not rest. Even in the absence of excitations, the magnetic field gradients must exert a force on the atoms which is greater then gravity for them to remain trapped. This limits the extent to which magnetic traps can be decompressed. Specially designed gravito-magnetic traps have been decompressed down to 1 Hz \cite{Leanhardt-03B} to investigate very cold, dilute BECs.

\section{Deep Trap Limitations}
\label{deeptraps}

A major difference we have observed between $^{87}$Rb and $^{23}$Na condensates is the
unexpectedly high decay rate of $^{87}$Rb in tightly confining deep traps, such as those 
used for transport in an optical trap \cite{Gustavson-02}. At the typical densities of condensates, 
the lifetime and heating are usually determined by three-body recombination decay.
However, the factor of four difference in the three-body rate coefficients (Table
\ref{tablerbprops}) was insufficient to explain this major
discrepancy in behavior.

We investigated this issue in a
magnetic trap instead of an optical trap.  While it is easier experimentally to create 
tight trapping and hence high densities in an optical trap,
both the trap frequencies and trap depth are functions of the
optical power. This makes it difficult to separate density dependent
effects, which are strongly affected by the trap frequency, from
trap depth effects in an optical trap. In contrast, in a magnetic trap the trap depth
can be controlled independently of the trap frequencies by adjusting the RF frequency
which flips atoms to untrapped states.

There are two possible processes, both involving secondary collisions, which
can greatly enhance the heating and losses due to the primary
three-body collisions.

The first process is collisional avalanches, similar to a chain
reaction, where the energetic products of three-body recombination
collide with further atoms while they leave the condensate.  This
process would depend on the collisional opacity $\sim n\sigma l$
(where $\sigma=8\pi a^2$ is the atom atom scattering cross
section) and would increase dramatically when it exceeds the critical opacity of 
0.693 \cite{Schuster-01}. This process is independent of trap depth.

The second possible process can already happen at lower collisional
opacities and relies on the retention of primary or secondary
collision products by the trap in the so-called Oort cloud \cite{Varenna,Burt-97}. Subsequently, when
those atoms slosh back into the trapped condensate, they can
cause heating and trap loss.  The retention of collision products in the Oort cloud should 
depend on whether the trap depth is larger or smaller than their energies.

\begin{figure}[ht!]
\begin{center}
\epsfig{file=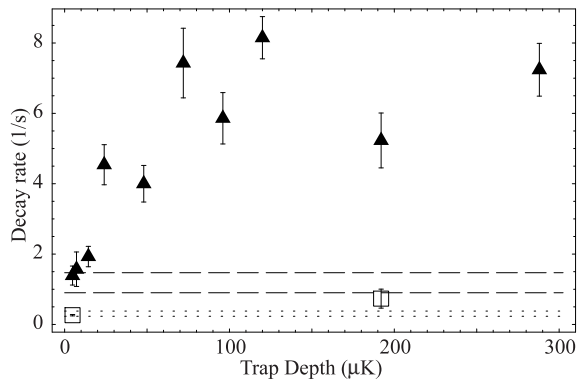, width=8 cm}
\caption{Initial loss rates for $^{87}$Rb BEC in deep traps. Trap depth dependence of the loss for large and small $^{87}$Rb BECs in a 220 Hz x 220 Hz x 9 Hz magnetic trap. The trap depth was controlled by RF truncation. Condensates were nearly pure ($N_c/N >$ 90\%) and consisted of F=1, m$_F$=-1 atoms. Solid triangles are data for a large condensate, $N_c=2.7\times10^6$ atoms, peak density $n_p=6.1\times10^{14}$/cm$^{3}$, expected three body decay time $\tau_3=0.85\pm0.22 $ s (dashed horizontal lines) \protect\cite{Tolra-03}, and calculated collisional opacity of 0.88 \protect\cite{Schuster-01}. Open squares are data for a small condensate, $N_c=5\times10^5$ atoms, peak density $n_p=3.1\times10^{14}$/cm$^{3}$, expected three body decay time $\tau_3=3.3\pm0.8$ s (dotted horizontal lines), and a calculated collisional opacity of 0.32. The error bars represent the statistical uncertainity in the decay curves. Additional scatter in the data is due to fluctutions in the atom number.}
\label{figrblossplot}
\end{center}
\end{figure}

Fig. \ref{figrblossplot} shows the initial loss rates measured for a large and
a small BEC as a function of the magnetic trap depth. At low trap depths (5 $\mu$K)
both the large and small condensate decay rates are in agreement with established three
body recombination rates \cite{Tolra-03}. Therefore, the avalanche effect does not significantly
contribute to the observed decay rate, although the calculated opacity for the larger condensate was 0.88 and may not be far away from the onset of avalanches. Evidence for avalanches was obtained at an opacity of 1.4 in  \cite{Schuster-01}.

In the larger condensate at 
higher trap depth, the decay rate strongly increases, supporting the second process involving the Oort cloud as the likely mechanism. For times longer then 500 ms, the large $^{87}$Rb condensate in the deeper traps was heated away and only a few thermal atoms remained. In contrast, at low trap depths the large condensate was still fully condensed after 20 seconds, and had an atom number in agreement with the expected losses from three body decay.

We speculate that this enhancement of the three-body losses was
not observed in $^{23}$Na for the following reasons. Three body
recombination results in a diatomic molecule and an atom which fly
apart with  a total kinetic energy (2/3 to the atom,
1/3 to the molecule to conserve momentum) equal to the binding
energy of the diatomic molecule in the highest vibrational state.
This binding energy can be estimated from the scattering length as
$E_0 \sim \hbar^2/ma^2$ \cite{Borca-03} ($\sim 200\mu$K in
$^{87}$Rb, $\sim$ 2.7 mK in $^{23}$Na). The direct decay products
will only be retained if the trap depth is greater then their
kinetic energies (min $\sim70\mu$K for $^{87}$Rb, $\sim900\mu$K
for $^{23}$Na).

In addition, $^{23}$Na decay products are less likely to undergo secondary collision
processes from either primary three body products or hot Oort cloud atoms due to
an elastic scattering cross section $\sigma$
which is 3.6 times smaller than for $^{87}$Rb. The individual products of secondary collisions
can have a spectrum of energies, further lowering the energy threshold for their retention by the Oort cloud. The combination of these
three factors, three body rate, scattering cross section, and
binding energy result in an estimated increase in the loss rate
for $^{87}$Rb condensates over $^{23}$Na by a factor of $\sim$200.

The optical trap depths needed for transporting condensates in our system are significant fraction of the primary $^{87}$Rb decay product energy, but a small fraction of that for $^{23}$Na. Therefore $^{23}$Na condensates can be easily transported using optical tweezers. For $^{87}$Rb the preferred method is to transport a cloud at temperatures just above condensation, where the density is lower, and evaporate to BEC after transport.

\section{Discussion}
\label{RbvsNa}

\renewcommand{\arraystretch}{1.3}
\begin{table}[h!]
\begin{tabular}[t]{|l|l|l|}
\hline
                                     & $^{87}$Rb & $^{23}$Na \\\hline\hline
D$_2$ line $\lambda$ (nm)            & 780                               & 589  \\\hline
D$_2$ linewidth $\Gamma/2\pi$ (MHz)  & 6.1                               & 9.8  \\\hline
Gravity $mg/k_B$ (nK/$\mu$m)         & 102                               & 27  \\\hline
Gravity $mg/g_F m_F \mu_B$ (G/cm)              & 30                                & 8.1 \\\hline
Three body constant $K_3$ (cm$^6$/s) & $8\times10^{-30}$ \cite{Tolra-03} & $2\times10^{-30}$ \cite{Gorlitz-03} \\\hline
Scattering length $a$ (nm)             & 5.3 \cite{Harber-02}              & 2.8 \cite{Samuelis-00} \\\hline
Recoil velocity $v_r$ (mm/s)         & 5.9                               & 29  \\\hline
\end{tabular}
\caption{ Select properties $^{87}$Rb and $^{23}$Na F=1, m$_F=-1$ ground states. Unless noted, quantities are derived from Ref. \protect\cite{Steck-02} and \protect\cite{Steck-03}. }
\label{tablerbprops}
\end{table}
\renewcommand{\arraystretch}{1}

$^{87}$Rb and $^{23}$Na are the two most popular species for
BEC research. We have constructed two machines with nearly
identical designs and can discuss differences in performance and
operation. Key properties of the two atoms are highlighted in
Table \ref{tablerbprops}. The four principal differences are in
vapor pressure, resonant wavelength, recoil velocity, 
and collisional properties.

The high vapor pressure of rubidium allows the operation of the
oven at lower temperatures, but requires a more elaborate design of
cold plates to avoid deposition of rubidium on surfaces of the UHV
chamber which are at room temperature. An optimized Zeeman slower for
rubidium will be about twice as long than that for sodium at similar oven temperatures, 
the stopping length L for the most probable velocity in a beam of temperature T being
$L= \frac{3k_B T}{\hbar k \Gamma}$, assuming the maximum spontaneous light force. In our systems
the gain from the greater light force is balanced out by the higher operating temperatures required of the sodium oven to produce comparable flux, resulting in both the rubidium and sodium slower being about 1 m in length.

Due to the higher recoil velocity, the slow sodium beam has a larger divergence than the
rubidium beam. By keeping the distance between the end of the
slower and the MOT to a minimum, we estimate 
quantitative transfer of atoms from the slower to the MOT.  Our
setup for rubidium was almost identical, but we expect that the
requirement of keeping the slower and the MOT so close could be
more relaxed for rubidium. Although we have not tried it, we
expect that our rubidium experiment would work for sodium
without changes to the oven, vacuum or magnet designs.

On the laser side, a major difference is the availability of low
cost high power laser diodes in the near infrared region around
780 nm. In our experience a well run dye laser system provides
similar or even superior performance to a diode laser system with
several master and slave lasers.  Our dye lasers at 589 nm tend to
be more stable than semiconductor lasers during a long run of the
experiment and need less day-to-day tweaking. However,
occasionally they require major maintenance in terms of dye
changes or full optical realignment. Another advantage is the
visibility of the laser light and the atomic fluorescence.  The
near infrared 780 nm light is only modestly visible, whereas the
sodium line at 589 nm is near the peak of human eye sensitivity
and allows fine alignments of the laser beams and the
magneto-optical trap without cameras, IR cards, or IR viewers.

$^{87}$Rb has favorable properties for laser cooling and atom
interferometry because of its greater mass, lower recoil velocity,
and larger excited state hyperfine structure. While greater mass
and longer resonant wavelength give $^{133}$Cs an even lower
recoil velocity, its complicated collisional behavior at low
magnetic fields makes it difficult to cool to BEC.
The lowest molasses temperature in rubidium is a factor of ten lower than for
sodium.
However, in BEC experiments the laser cooling is
optimized to for large atom numbers and high initial elastic collision rates in the magnetic trap,
and not for the lowest temperature.  For laser cooling sodium at
high atom numbers, the Dark SPOT technique \cite{Ketterle-93} is crucial to avoid
rescattering of light, whereas it is only used in some rubidium
experiments. One possible reason is that the larger excited state
hyperfine structure allows for larger detunings from the cycling
transition without exciting other hyperfine states. At the end of
the day, although with somewhat different techniques, the laser
cooling part works equally well for both atoms.

Both atomic species have favorable collisional properties for
evaporative cooling.  The elastic scattering cross section of
$^{87}$Rb atoms at low temperature is four times higher than in
$^{23}$Na.  However, elastic collision rates after laser cooling are
comparable since $^{23}$Na atoms are faster.  A peculiarity of
$^{87}$Rb is that the two ground electronic state hyperfine levels have
similar scattering lengths, which can be advantageous for studies
on spinor condensates and atomic clock transitions. Related to
that, spin relaxation between the two hyperfine levels is almost
completely suppressed. Mixtures of F=1 and F=2 atoms can be kept for seconds \cite{Harber-02}, whereas in $^{23}$Na they decay on ms time scales \cite{Gorlitz-03}.
Both atoms have several Feshbach resonances
below 1100 G \cite{Inouye-98,Stenger-99b,Marte-02}. Here, $^{87}$Rb
has the disadvantage, that the widest known resonance is only 200
mG wide compared to 1 G for $^{23}$Na and requires more stable
magnetic fields.  Another difference is the higher rate of
three-body collisions for $^{87}$Rb atoms.  As we discussed in Sec.
\ref{deeptraps}, this imposes limitations on trapping and manipulating dense $^{87}$Rb
condensates.

%
%
%
%
\section{Conclusions}
\label{Acknowledgments}


In this paper, we have presented details for designing
BEC machines with high performance and flexibility, and we hope
that this description is useful for designing new experiments.
Given the recent developments in the field, there is more than
enough room for new experiments to join in the exploration of atom
optics and many-body physics with quantum-degenerate atomic gases.

Funding for the $^{87}$Rb machine was provided by the NSF
MIT-Harvard Center for Ultracold Atoms. Funding for the $^{23}$Na
machine came from NSF, the ARO MURI program, NASA, and the ONR. 

We thank S. Gupta,  A. G{\"o}rlitz, and A. E. Leanhardt for their contributions in the construction 
of the $^{23}$Na machine; J. C. Mun and P. Medley for ongoing contributions to the $^{87}$Rb machine; and MIT UROP students P. Gorelik and X. Sun for various contributions to the $^{87}$Rb machine. The authors would 
also like to thank M. Saba and D. Kielpinksi for critical reading of this manuscript.

%
%
%
%

\appendix
\section{Oven}
\label{ovenappendix}
To sustain a high flux atomic beam, the background vacuum pressure must be low enough that the mean free path between collisions is much greater than the length of the beam. To generate an effusive beam with a thermal distribution of velocities, the size of the hole through which the atoms escape must be smaller than the mean free path inside the oven. We observed in sodium that at higher pressures (e.g. temperatures) the flux of
slowable atoms does not increase and that the velocity distribution narrows. This phenomena is well understood \cite{AtomicMolecularBeamMethods}, and limits the diffusive flux from a single aperture oven.

\begin{table}
\begin{minipage}[b]{8cm}
\begin{tabular}[b]{|l|l|l|l|l|}
\hline
Temp & Velocity\footnote{Most probable,3D Beam, Sec 5.2 Ref. \cite{Metcalf}} & Pressure \cite{Alcock-84} & Total Flux\footnote{5mm aperture} & Lifetime\footnote{5g Rb}\\
($^\circ$C) & (m/s) & (torr) & \#/sec & (hours) \\\hline\hline
-30&266&2.5E-10&1.4E+10&--\\\hline
25&295&4.0E-07&2.0E+13&--\\\hline
40 \footnote{melting point} &302&2.0E-06&9.7E+13&--\\\hline
90&325&1.2E-04&5.4E+15&1815\\\hline
100&330&2.3E-04&1.1E+16&926\\\hline
110&334&4.5E-04&2.0E+16&489\\\hline
120&339&8.3E-04&3.7E+16&267\\\hline
130&343&1.5E-03&6.5E+16&151\\\hline
140&347&2.6E-03&1.1E+17&87\\\hline
150&351&4.4E-03&1.7E+17&52\\\hline
\end{tabular}
\caption{Design Parameter for Rb oven}
\label{tablerbdesign}
\end{minipage}
\end{table}

During servicing, a clean ampoule is essential for rapid recovery of good vacuum pressure. The ampoule is cleaned by submerging it in a 50/50 mixture by volume of acetone and isopropanol for 20 minutes, air drying it, and then by placing it in the oven while still sealed. This removes most of the water from the glass surface, which would otherwise require more time to pump away. After installation the ampoule is baked for 24 hours under vacuum at 150-180$^\circ$C to remove the remaining contaminates before it is broken.

To prevent accumulation of metal at the aperture (Fig. \ref{figovendiagram}, J), the oven nozzle temperature (Fig. \ref{figovendiagram}, K) is kept hotter ($\sim10^\circ$C in rubidium and $\sim90^\circ$C in sodium) than the rest of the oven . The velocity distribution of the beam is determined by the nozzle temperature (Fig. \ref{figovendiagram}, K). On the other hand, the vapor pressure in the oven, which controls the beam flux, is dominated by the coldest spot in the elbow and bellows. The factor of two discrepancy between the observed and calculated (Table \ref{tablerbdesign}) rubidium oven lifetimes at 110$^\circ$C can be accounted for by a spot $\sim10^\circ$C colder then the lowest measured oven temperature. The specifics of this cold spot depend on how the oven is insulated.

%
%
%
%

\section{Zeeman Slower}
\label{slowerappendix}

Every photon which scatters off an atom to slow the atom is radiated in a random direction, increasing the atoms spread in transverse velocity. The beam emerging from the tube needs to have sufficient forward mean velocity to load the MOT efficiently. Because of the random direction of the emission recoil, N photon scatterings increase the transverse velocity by $v_r \sqrt{N/3}$, or $\sqrt{v_r \Delta v/3}$. The $^{23}$Na slower operates with a recoil induced transverse exit velocity of $\approx3$m/s, with a final forward velocity of 30 m/s so that the spatial transverse spread in the slowed beam matches the MOT capture area. The smaller initial and recoil velocities in the $^{87}$Rb slower reduce the transverse velocity to $\approx0.8$m/s, making MOT capture matching less critical.

An additional concern in both slowers is the fate of atoms not captured by the MOT. In $^{87}$Rb we were concerned with the potential adverse impact a deposited film may have on the vapor pressure, and installed a cold plate near the slower window on the main chamber to capture desorbed Rb. Vacuum pressure has not been an issue and we have never needed to chill this cold plate. The opposite problem arises in $^{23}$Na, where metal deposition on the slower window reduces the transmission of slower light. We have found heating the slowing beam vacuum port window to 90$^\circ$C prevents long term buildup.

\subsection{Slower Construction}

The vacuum portion of the $^{87}$Rb slower is a 99 cm long nonmagnetic 304 stainless steel tube with a 19 mm OD and 0.9 mm wall. The rear end of the tube is connected to the main chamber by a DN 16 CF rotatable flange, while the oven end of the tube has a narrow, 50mm long flexible welded bellows ending in another DN 16 CF rotatable flange. The retaining ring on this flange was cut in half for removal, so that the premounted coil assembly could be slid over the vacuum tube.

As shown in Fig. \ref{figwholesystem} the slower tube enters the main chamber at an angle of $33^\circ$ from the vertical to accommodate access for optical tweezers. The oven and the Zeeman slower are supported two meters above the the experimental table in order to preserve the best
optical and mechanical access to the main chamber. Aluminum extrusion from 80/20 Inc. was used to create the support framework.

\begin{figure}[ht!]
\begin{center}
\epsfig{file=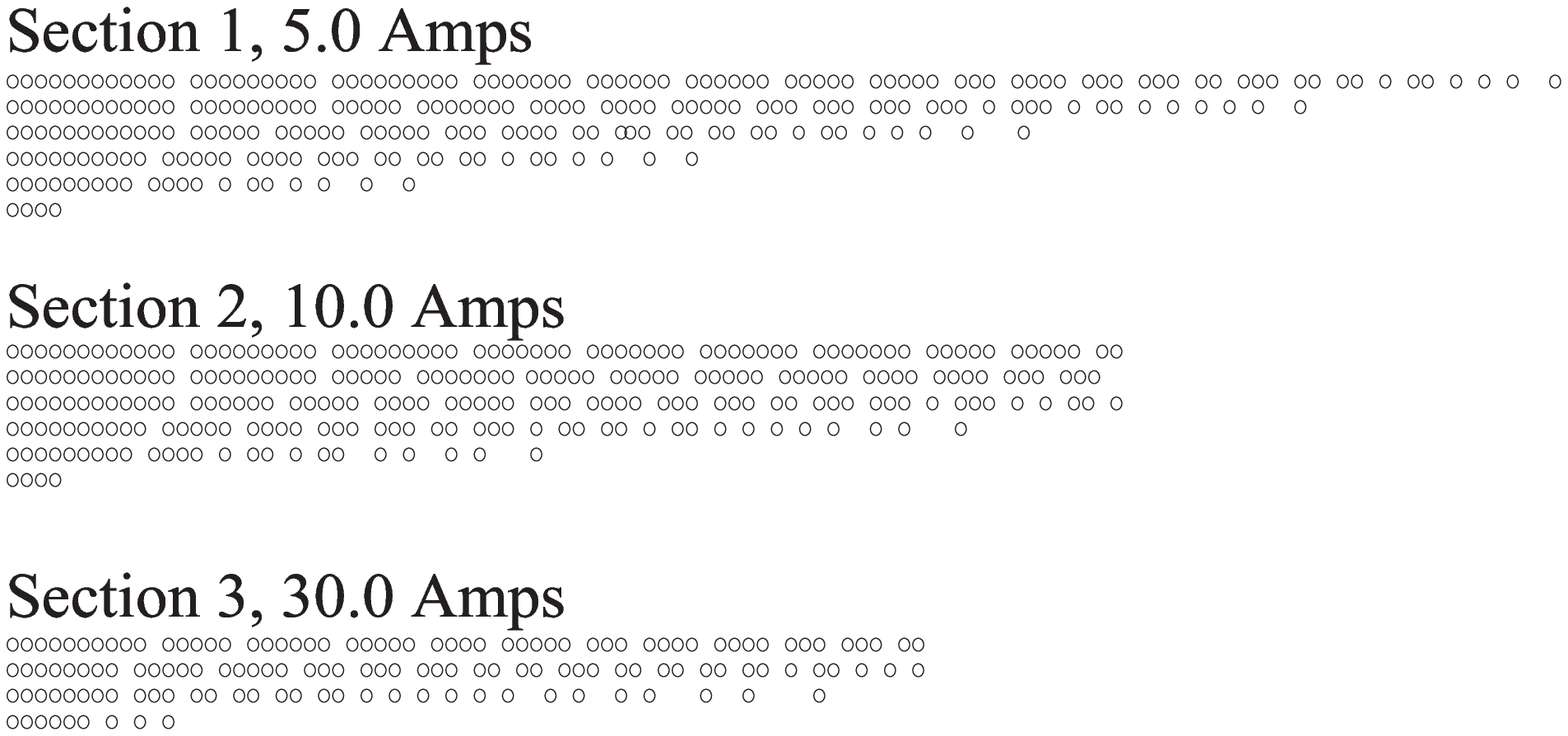, width=8 cm}
\caption{Winding pattern cross section for $^{87}$Rb Zeeman slower consisting of three solenoids.  Each drawing represents half of the cross section of each a solenoid.  The ``O''s represent wires, while the spaces between the wires were meant to be smoothed out to an average value during construction.  Each character in the drawing represents a physical size of 3.5 mm. The wire is hollow core water cooled copper, identical to that used in construction of the magnetic trap as described in \ref{HollowCoreWire}. The high current coil is closest to the main chamber. The single layer uniform bias coil is not depicted.}
\label{figslowerwinding}
\end{center}
\end{figure}

Our $^{87}$Rb slower was fabricated with a single layer bias solenoid and three increasing field segments (Fig. \ref{figwholesystem} and \ref{figslowerwinding}). The optimum configuration of currents and solenoid winding shapes was found by computer simulated winding of the solenoids one loop at a time, starting at the high field end and tapering the last few loops to best match the desired field profile. An alternative fabrication technique would be to apply a large uniform bias field and subtract away unwanted field with counter current coils. Stray fields from the Zeeman slower can have a detrimental effect on the MOT, particularly during sudden turnoff.  An additional bias coil around the main chamber along the axis of the slower can compensate for this effect.

\section{Magnetic Trap Appendix}
\label{appendixMagtrap}

\subsection{Ioffe Pritchard Trapping Potential}
The field near the minimum of a Ioffe-Pritchard trap is approximately
\begin{equation}
\mathbf{B} = B_0\left(\begin{array}{c} 0\\0\\1\\ \end{array} \right)   + B'\left(\begin{array}{c} x \\ -y \\ 0 \\ \end{array} \right) + \frac{B''}{2}\left(\begin{array}{c} -xz \\ -yz \\ z^2- \frac{1}{2} \left(x^2+y^2\right)\\ \end{array} \right)
\label{eqnIPmagneticfield}
\end{equation}

which realizes trap frequencies

\begin{eqnarray}
m\omega_{\perp}^2 = & \mu_m \frac{B'^2}{B_0} \\
m\omega_{z}^2  = & \mu_m B''
\label{eqnIPtrapfrequencies}
\end{eqnarray}

Typical trap parameters of $B^\prime$=223 G/cm, $B^{\prime\prime}$=100 G/cm$^2$, $B_0$=1G (Fig. \ref{figiptrapprofile}, Table \ref{tablecoilvalues}) have frequencies of $\left( \omega_{\perp}, \omega_{z} \right)/2\pi$ of $(200 , 9 )$ Hz for $^{87}$Rb and $(390, 18)$ Hz for $^{23}$Na. Further details of Ioffe-Pritchard magnetic traps are discussed in Sec 2.3.2 of \cite{Varenna} and Ch. 5. of \cite{Durfee-99}.

\subsection{Circuitry}

\begin{figure}[ht!]
\begin{center}
\epsfig{file=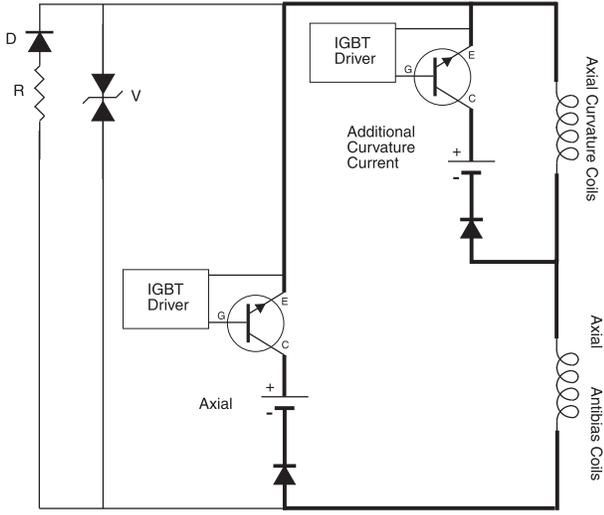, width=8 cm}
\caption{Axial coils circuit diagram. High current wires are heavy black lines.}
\label{figaxialcircuit}
\end{center}
\end{figure}

Fig. \ref{figaxialcircuit} is representative of the magnetic trap circuit. We drive the magnetic trap coils with Lambda EMI DC power supplies in fixed current mode. Current to the cloverleaf coils is supplied from a Model ESS 30-500 15kW power supply, while the axial currents are driven with two Model EMS 20-250 5kW power supplies. Each power supply is protected against damage from reverse current with an International Rectifier SD600N04PC high-current diode.

To switch the high currents we use PowerEx models CM1000HA-24H and CM600HA-24H Integrated Gate Bipolar Transistors (IGBTs) controlled with PowerEx BG1A-F IGBT driver kits. The IGBTs and high current diodes dissipate several hundred W during operation, and are cooled with chilled water. Efficient heat sinking is critical for reliable operation, as thermal dissipation limits the maximum DC current. Fast turnoff of current on an inductive load, such as a coil, results in a large voltage spike. We have added a ``debounce'' circuit to each of the coil systems (Fig. \ref{figaxialcircuit}) to control this process and prevent damage. The circuit consists of two different elements: a varistor (V) and a diode (D) in series with a low impedance resistor (R). The varistor shorts the circuit at high voltages to prevent this spike, and the diode (D) in series with a 1$\Omega$ resistor (R) dissipates the remaining current after varistor shutoff. Ref. \cite{Melton-96} contains a through analysis
of the behavior of a similar circuit. All control signals are electrically isolated from the high-current circuits to prevent voltage spikes from damaging connected hardware. Rapid, controlled magnetic field shutoff is important for quantitative interpretation of images taken after ballistic expansion.

\subsection{Wire}
\label{HollowCoreWire}
Both the slower and the magnetic trap coils were fabricated using square hollow core (0.125 in./side, 0.032 in. wall) Alloy 101 soft temper copper tubing from Small Tube Products, Inc. of Altoona, PA, wrapped with double Dacron glass fuse insulation by Essex Group Inc., Magnet Wire \& Insulation of Charlotte, NC. The coils are held together with Hysol Epoxi-Patch 1C White high temperature epoxy that is bakable to 170$^\circ$C. Chilled water is forced through the hollow core of the copper wires to dissipate the $\approx$10kW of power generated from resistive heating in the magnetic trap and Zeeman slower coils. 200 psi of differential pressure is required for sufficient coolant flow. We designed all our coils to increase the cooling water temperature by less then 50$^\circ$C. Ch. 3 of Ref. \cite{SolenoidDesign} has an extensive discussion of water cooling in continuously powered resistive magnets. For our wire the following empirical values were measured,

\begin{eqnarray}
\rho \left[ \Omega/ m \right] & = & 2.65\times10^{-3}   \\
Q \left[ ml/sec \right] & = & 2.07  \sqrt{\frac{\Delta P \left[ psi \right] }{L \left[ m \right]}}  \\
\Delta T \left[ ^\circ C \right] & = &  259   I^2 \left[ Amps \right] \rho \sqrt{ \frac{L^3 \left[ m \right]}{\Delta P \left[ psi \right]}}
\label{eqnssquarewalltubing}
\end{eqnarray}

where Q is the water flow rate in ml/sec, I$^2$ $\rho$ L is the power dissipated by the coil, $\Delta$P the pressure drop in psi (1 psi=6.89 kPa), and L the length of the coil in meters.

\subsection{Fabrication}
All of the components for each half of the magnetic trap were epoxied together for stability. Each assembly was then mounted in the bucket windows with an aluminum mounting plate backed by four threaded Alloy 316 stainless steel rods. No ferromagnetic materials were used in the mounting because of concern for irreproducibility from hysteresis effects.  Table \ref{tablecoilvalues} lists the windings and typical parameters for each coil.

\renewcommand{\arraystretch}{1.2}
\begin{table}[h!]
\begin{minipage}[t]{8cm}
\begin{tabular}[t]{|l|ll|l|l|l|}
\hline
{\bf Coil} & {\bf Winding} &        & {\bf Current} & {\bf Inner $\emptyset$} & {\bf Field} \\
           & Turns         & Layers & (A)           & (cm)                    &  \\\hline
Antibias  & 3 & 6 & 95  & 10.5 & $B''$ = +9 G/cm$^{2}$\\
          &   &   &     &      & $B_0$ = -243 G \\\hline
Curvature & 8 & 6 & 95  & 3.2 & $B''$ = +90 G/cm$^{2}$\\
          &   &   &     &      & $B_0$ = +251 G \\\hline
Gradient  & 3 & 4 & 470 & 0.8 x 2.3 & $B'$ = 223 G/cm\\\hline
MOT       & 7 & 4+2 \footnote{Segmented for improved cooling.} & 15 & $\sim$ 7, to fit & 16.5 G/cm in $\hat{z}$\\\hline
\end{tabular}
\caption{Magnetic trap coil winding and performance specifications. Fig. \ref{figexplodedcoils} illustrates their assembly and direction of current flow.}
\label{tablecoilvalues}
\end{minipage}
\end{table}
\renewcommand{\arraystretch}{1.0}

%
%
%
%

%
%
%
%

\end{document}